\documentclass[12pt]{article}

\usepackage{graphicx}
\usepackage{cite}

\def\be{\begin{equation}}
\def\ee{\end{equation}}
\def\ba{\begin{eqnarray}}
\def\ea{\end{eqnarray}}
\def\la{~\mbox{\raisebox{-.6ex}{$\stackrel{<}{\sim}$}}~}
\def\ga{~\mbox{\raisebox{-.6ex}{$\stackrel{>}{\sim}$}}~}
\def\bq{\begin{quote}}
\def\eq{\end{quote}}

 at 10truept

\newcommand{\beq}{\begin{equation}}
\newcommand{\eeq}{\end{equation}}
\newcommand{\beqa}{\begin{eqnarray}}
\newcommand{\eeqa}{\end{eqnarray}}

\def\la{~\mbox{\raisebox{-.6ex}{$\stackrel{<}{\sim}$}}~}
\def\ga{~\mbox{\raisebox{-.6ex}{$\stackrel{>}{\sim}$}}~}

\def\ltap{\ \raise.3ex\hbox{$<$\kern-.75em\lower1ex\hbox{$\sim$}}\ }
\def\gtap{\ \raise.3ex\hbox{$>$\kern-.75em\lower1ex\hbox{$\sim$}}\ }
\def\gl{\ \raise.5ex\hbox{$>$}\kern-.8em\lower.5ex\hbox{$<$}\ }
\def\roughly#1{\raise.3ex\hbox{$#1$\kern-.75em\lower1ex\hbox{$\sim$}}}

\begin{document}
\thispagestyle{empty}
\begin{flushright}
{\tt hep-ph/0302030} \\ CITA-2003-03
\end{flushright}
\vspace*{0.2cm}
\begin{center}
{\Large \bf Super-GZK Photons from Photon-Axion Mixing}\\
\vspace*{1cm}
{\large Csaba Cs\'aki$^{a}$,
Nemanja Kaloper$^{b}$,
Marco Peloso$^{c}$
and John Terning$^{d}$}

\vspace{0.5cm}

{\em $^a$ Newman Laboratory of Elementary Particle Physics}\\
{\em Cornell University, Ithaca, NY 14853, USA}

\vspace{0.2cm}

{\em $^b$ Department of Physics, University of California}\\
{\em Davis, CA 95616, USA}

\vspace{0.2cm}

{\em $^c$ Canadian Institute for Theoretical Astrophysics, University of Toronto}\\
{\em 60 St.~George St., Toronto, ON M5S 3H8, Canada}
\vspace{0.2cm}

{\em $^d$ Theory Division T-8, Los Alamos National Laboratory}\\
{\em Los Alamos, NM 87545 USA}\\
\vspace{0.2cm}
{\tt csaki@mail.lns.cornell.edu, kaloper@physics.ucdavis.edu,
peloso@cita.utoronto.ca, terning@lanl.gov}

\vspace{1cm}
ABSTRACT
\end{center}
We show that photons with energies above the GZK cutoff can reach us
from very distant sources if they mix with light axions in extragalactic magnetic fields.
The effect which enables this is the conversion of photons into axions, which are
sufficiently weakly coupled to travel large distances unimpeded. These axions
then convert back into high energy photons close to the Earth. We show that
photon-axion mixing facilitates the survival of super-GZK photons
most efficiently with a photon-axion coupling scale $M \ga 10^{11}$ GeV,
which is in the same range as the scale for the photon-axion mixing explanation
for the dimming of supernovae without cosmic acceleration.
We discuss possible observational consequences of this effect.

\vfill
\setcounter{page}{0}
\newpage

\section{Introduction}
\setcounter{footnote}{0}

Observations suggest that
Ultra High Energy Cosmic Rays (UHECRs) with
energies $\sim {\rm few} \times 10^{20} \,$ eV may exist in Nature
(for reviews, see e.g.
\cite{bhsi,nawa,aprs}). Usually it is assumed that such particles are
of astrophysical origin.\footnote{Alternatively, it
has been suggested that they may originate from the nearby decay of very
massive particles \cite{kuru,bkv} or topological defects \cite{hill}.}
While different candidate sources of UHECRs have been proposed, there is a
consensus that they should be at distances farther than ${\cal O}(100)\,$ Mpc.
Such distances are difficult to reconcile with the observed energies of UHECRs
in simple scenarios. Indeed, if the
UHECRs are protons, any flux above $\sim 5 \times 10^{19} \,$ eV --
the well-known Greisen-Zatsepin-Kuzmin (GZK) cutoff \cite{gzk} --
is strongly attenuated by the photo-pion production on the Cosmic Microwave
Background (CMB) photons. If the UHECRs are photons, their flux is in turn
strongly attenuated by the scattering against the
extragalactic radio background, seemingly ruling them out as the candidates
for the highest energy UHECRs \cite{prjo}.

A way around the GZK cutoff might be a mechanism where the cosmic
rays mostly consist of some very weakly interacting particles,
which could travel long distances without being significantly
absorbed. The only experimentally observed particles which could
do this are neutrinos. A few mechanisms with neutrinos as the
mediators, traveling most of the distance between a source and us
have been suggested. Such proposals, however, suffer from the
``converse" problem: how does a sufficient fraction of the neutrinos
convert into nucleons or photons near the Earth. It has been
argued \cite{weiler,roulet,fargionweiler,ysl,fkr}
that, for neutrino masses of order eV, the
UHE neutrinos can interact significantly with the cosmological
background neutrinos through the $Z$ boson resonance. This
mechanism however may require very high incoming fluxes and a very
strong local clustering of the background neutrinos to conform
with the observations \cite{ysl}.

It is therefore natural to ask if other particles (from
beyond the standard model (SM) of particle physics) could be such mediators.
Recently, pseudoscalars arising from the spontaneous breaking
of an axial symmetry, or axions for short (similar to  the QCD-axion
\cite{axion}), have been considered in an alternative explanation for
the apparent dimming of distant type Ia supernovae (SNe) \cite{ckt}. The idea was
that a fraction of the photons emitted by a supernova could convert
into axions in the presence of an intergalactic magnetic field
en route to the Earth, thereby dimming the supernova and making it look farther
away than it is \cite{ckt}. Since these axions can travel
very far without significant interactions, and they can convert to photons (and vice-versa)
in background magnetic fields  (see, e.g. \cite{osc}),
they are a natural candidate for a
mediator which could help evade the GZK cutoff.

Ref. \cite{grs} investigated whether the QCD-axion, which could arise from a
solution of the strong CP problem \cite{axion}, could act as a
mediator of UHECRs. Because the photon-axion
coupling is of the form
\begin{equation}
\label{coupling}
\frac{a}{4 M} {\tilde F} \, F\ ,
\end{equation}
where $a$ is the axion field, and $F$ the electromagnetic field strength, the
evolution in the presence of a background magnetic field $B$ induces
``flavor" (photon-axion)
conversions. The trilinear interaction (\ref{coupling}) gives rise to a
bilinear term which is not diagonal in ``flavor" space.
The probability for the photon-axion oscillation in a homogeneous magnetic field is
\begin{eqnarray}
P_{a \rightarrow \gamma} &=& \frac{1}{1+\frac{\Delta m^4 M^2}{4 \, {\cal E}^2 \,
B^2}} \, {\rm sin}^2 \left[ \frac{y}{2} \, \sqrt{
\frac{B^2}{M^2} + \frac{\Delta m^4}{4 \, {\cal E}^2}} \right] \,\,,
\label{prob} \\
\Delta m^2 &\equiv& m_\gamma^2 - m_a^2
\label{Deltam2}
\end{eqnarray}
where $m_\gamma$ is the effective photon mass in the intergalactic
plasma, $m_a$ is the axion mass, ${\cal E}$ the energy of the
particles, and $y$ the path length. For a QCD axion, $10^{-5} \,
{\rm eV} \la m_{a, {\rm QCD}} \la 10^{-1} \,$ eV, so the
conversion $a \rightarrow \gamma$ is completely negligible
\cite{grs}. One way around this problem pursued in~\cite{grs} was
to assume that the axion does not convert into photons, but that it
interacts with the Earth's atmosphere due to its
relatively large coupling to gluons. Here we
will follow a different direction. In the photon-axion model for
the dimming of the SNe a much lighter axion is invoked \cite{ckt}
(see also \cite{otherckt}). Then the axion mass will not suppress
the mixing even for the optical photons contributing to the SNe
luminosity, let alone for the UHECRs with energies larger by
nearly 20 orders of magnitude. This observation is the starting
point of this paper, where we will show that the coupled
photon-axion system could indeed give rise to the highest energy
cosmic rays. We will also compare the preferred region of the
parameter space where the transmission of ultra-high energy
photons is maximal to the parameters needed to explain the dimming
of the SNe, and show that the UHECR flux due to the photon-axion
mixing is especially enhanced for the choice of the parameters
needed to account for the SNe dimming in \cite{ckt}. Since the
mean free path $l_{\rm GZK}$ of the photons is very sensitive to
their energy, and since the energy of the UHECRs is nearly 20
orders of magnitude higher than the ones of the SNe, this
coincidence is very intriguing.

Our model for the transmission of UHECRs works as follows.
Initially, a faraway astrophysical source releases a flux of very
high energy unpolarized photons. A small fraction of these photons
converts into axions before being depleted by the interactions
with the background (radio and CMB) photons\footnote{The photons
which do interact with the background photons give rise to
secondary photons which can also propagate along the beam with a
lower energy than the primary ones. To simplify the present
discussion we will neglect those and concentrate on particles
which have never interacted with the background photons.}. At
distances $d \gg l_{\rm GZK}$, defined as the mean free path of
the photons in absence of the mixing with axions, the beam is
mostly comprised of axions\footnote{Alternatively, one can start
with only axions from the very beginning, as in ref.~\cite{grs},
assuming that they are produced directly at the source.}. These
axions will gradually convert back into photons, but most of these
photons will be depleted by the interaction with background
photons. However, the conversion rate per distance traveled is
low, allowing most of the axions to travel unimpeded, and keep
replenishing the photon beam. This will ensure that a fraction of
the photons will survive over distances much larger than one would
expect from the GZK length of UHE photons.

The paper is organized as follows: in Section 2 we describe the
evolution of the photon-axion system in the presence of the GZK
cutoff, making the (unrealistic) assumption of a homogeneous
magnetic field of order $\sim 10^{-9}$ Gauss. In this case the
evolution of the system can be calculated exactly, and we show
that there indeed exists a long-lived component in the photon
beam. In Section 3 we examine a more realistic situation, where
the magnetic field consists of domains of size $\sim$ Mpc, with a
random orientation of the field inside each domain, again with a
strength $\sim 10^{-9}$ Gauss \cite{magn}. We calculate the
surviving intensity of the photon beam as a function of the
distance traveled and the parameters of the photon-axion system,
and show that the highest survival probabilities are expected for
the same parameters that are preferred for the SNe dimming
mechanism of \cite{ckt}. We also discuss possible peculiar
signatures of this UHECR mediation mechanism arising from the fact
that most of the observed photons would be generated from a
predominantly axion beam within the last few magnetic domains
around the Earth. We estimate the initial photon flux needed for
this mechanism to work, and finally conclude in Section 4.

\section{Evolution of the photon-axion system}
\setcounter{footnote}{0}

The evolution of the photon-axion system propagating along the $y$
direction is governed by
\be
\Bigl\{ \frac{d^2}{dy^2} + {\cal E}^2
+ \pmatrix{ i \Gamma({\cal E}) & -i {\cal E} \frac{B}{M} \cr i
{\cal E} \frac{B}{M} & 0 \cr} \Bigr\} \pmatrix{ |\gamma\rangle \cr
|a\rangle \cr} = 0 \label{frice}
\ee
where we have Fourier-transformed the fields to the energy (${\cal E}$)
picture. Here $\vec B$ is the extra-galactic magnetic field and
$B =  \vec e \cdot \vec B \sim | \vec B|$ is its projection on the
photon polarization $\vec e$. In a constant magnetic field
$|\gamma\rangle$ denotes the photon polarization component along
$\vec B$, $|a\rangle$ the axion, while the photon polarization
$|\gamma_\perp\rangle$ orthogonal to $\vec B$ decouples from
$\vert a \rangle$ and evolves independently. In the next section
we will discuss the evolution of the photon-axion beam through a
sequence of domains with a different orientation of the magnetic
field in each domain, where both photon polarizations must be
retained. In both cases we take $|\vec B| \sim {\rm few} \cdot
10^{-9}$ G. Throughout our analysis we can ignore the axion mass
and the effective photon mass, since they are much
smaller\footnote{We assume that $m_a^2 \ll {\cal E} \, B / M$;
this ensures that the photon-axion oscillations are unsuppressed,
but also excludes the QCD axion.} than the beam energy ${\cal E}
\sim 10^{20} eV$, and $ {\cal E} \, B / M$. Instead we include the
(energy dependent) decay rate $\Gamma$, which parameterizes the
decrease of the photon intensity due to their interaction with the
background photons. If we rewrite $\Gamma \sim {\cal E}/l_{\rm
GZK}$, in the absence of the coupling to the axions $l_{\rm GZK}$
gives the mean free path of the photons. For ${\cal E} = 3 \cdot
10^{20} eV$ (which we will consider in our numerical examples
later on), $l_{\rm GZK} \simeq 6.4 \,$ Mpc~\cite{prjo}.

For a constant $B$, the solution of Eqs. (\ref{frice}) is
\begin{eqnarray}
\left( \begin{array}{c}
|\gamma\rangle \\ |a\rangle \end{array} \right) &=&
\left( \begin{array}{cc}
\frac{1+\sqrt{1-4 \delta^2}}{2\,\sqrt{1-4\,\delta^2}} &
-\,\frac{\delta}{\sqrt{1-4\,\delta^2}} \\
\frac{\delta}{\sqrt{1-4\,\delta^2}} &
-\,\frac{1-\sqrt{1-4 \delta^2}}{2\,\sqrt{1-4\,\delta^2}} \end{array} \right)
\, {\rm e}^{i\,{\cal E}\,y+\lambda_1 \, y} \,
\left( \begin{array}{c}
|\gamma\rangle \\ |a\rangle \end{array} \right)_0 + \nonumber\\
&+& \left( \begin{array}{cc}
-\,\frac{1-\sqrt{1-4 \delta^2}}{2\,\sqrt{1-4\,\delta^2}} &
\frac{\delta}{\sqrt{1-4\,\delta^2}} \\
-\,\frac{\delta}{\sqrt{1-4\,\delta^2}} &
\frac{1+\sqrt{1-4 \delta^2}}{2\,\sqrt{1-4\,\delta^2}} \end{array} \right)
\, {\rm e}^{i\,{\cal E}\,y+\lambda_2 \, y} \,
\left( \begin{array}{c}
|\gamma\rangle \\ |a\rangle \end{array} \right)_0  \,\,,
\label{sol2}
\end{eqnarray}
where the subscript zero denotes the initial amplitudes at the source at $y=0$,
and where we use
\begin{eqnarray}
\lambda_1 &\equiv& - \, \frac{1}{4\,l_{\rm GZK}} \, \left[ 1 +
\sqrt{1-4\,\delta^2} \right] \,\,,\nonumber\\
\lambda_2 &\equiv& - \, \frac{1}{4\,l_{\rm GZK}} \, \left[ 1 -
\sqrt{1-4\,\delta^2} \right] \,\,,\nonumber\\
\delta &\equiv& \frac{B\,l_{\rm GZK}}{M} \simeq 0.11 \, \left( \frac{B}{1
\,{\rm nG}} \right) \, \left( \frac{10^{11} \, {\rm GeV}}{M} \right) \,
\left( \frac{l_{\rm GZK}}{\rm Mpc} \right) \,\,. \label{deltaref}
\end{eqnarray}
Notice that (\ref{sol2}) is regular for all values of $\delta$.
One can see this by rewriting (\ref{sol2}) as
\begin{eqnarray}
&&\left( \begin{array}{c}
|\gamma\rangle \\ |a\rangle
\end{array} \right) = {\rm e}^{i {\cal E} y} \,
{\rm e}^{-\frac{y}{4 l_{\rm GZK}}} \,
\left( \begin{array}{cc}
{\cal C} - {\cal S} & 2 \, \delta {\cal S} \\
- 2 \, \delta {\cal S} & {\cal C} + {\cal S}
\end{array} \right) \,
\left( \begin{array}{c}
|\gamma\rangle \\ |a\rangle
\end{array} \right)_0 \nonumber\\
&& {\cal C} \equiv {\rm cosh } \left[ \sqrt{1-4 \delta^2} \frac{y}{4
l_{\rm GZK}} \right] \;\;,\;\;
{\cal S} \equiv \frac{{\rm sinh } \left[ \sqrt{1-4 \delta^2} \frac{y}{4
l_{\rm GZK}} \right]}{\sqrt{1-4 \delta^2}} \,\,,
\end{eqnarray}
with ${\cal C}$ and ${\cal S}$ real numbers for any choice of
$\delta\,$.

We are interested
in the mean free path of the photons in the presence of the coupling to the axions.
The initial photon beam is taken to be composed of unpolarized photons, that is it contains
an equal mixture of $|\gamma\rangle$ and $|\gamma_\perp\rangle$,
with the total intensity normalized to unity. Then the intensities of the
surviving photons and axions after the distance $y$ away from the source,
using eq. (\ref{sol2}) $I_\gamma$ and $I_a$ are
\begin{eqnarray}
I_\gamma &=& \frac{1}{2} \left\{ {\rm e}^{-\frac{y}{l_{\rm GZK}}} +
\frac{1}{4\,\left( 1-4 \delta^2 \right)} \, \left[ \left(1+\sqrt{1-4
\delta^2}\right) \, {\rm e}^{-\frac{1+\sqrt{1-4 \delta^2}}{4\,l_{\rm GZK}}y}
\right. \right.
\nonumber \\
&&
\left. \left.
- \left(1-\sqrt{1-4 \delta^2}\right) \, {\rm e}^{-\frac{1-\sqrt{1-4
\delta^2}}{4\,l_{\rm GZK}}y} \right]^2\right\} \, , \nonumber\\
I_a &=& \frac{\delta^2}{2\left(1-4 \delta^2\right)} \, \left[
{\rm e}^{-\frac{1+\sqrt{1-4 \delta^2}}{4\,l_{\rm GZK}}y}-{\rm e}^{-\frac{1
-\sqrt{1-4 \delta^2}}{4\,l_{\rm GZK}}y} \right]^2 \,\,.
\label{ana1}
\end{eqnarray}
The first term in (\ref{ana1}) gives the intensity of the photons orthogonal
to $\vec B$, which do not mix with the axion and thus simply decay
away as is expected from the GZK mechanism,
while the remaining terms give the intensities of the coupled
axion-photon system. They have an easy interpretation both in the limit
of large and small $\delta$. For a large mixing ($\delta \gg 1$),
\begin{eqnarray}
I_\gamma  &\simeq& \frac{1}{2} \, {\rm e}^{-\frac{y}{l_{\rm GZK}}} +
\frac{1}{2} \, {\rm e}^{-\frac{y}{2\,l_{\rm GZK}}} \, {\rm cos}^2 \left(
\frac{\delta y}{2\,l_{\rm GZK}} \right) \nonumber\\
I_a &\simeq& \frac{1}{2} \, {\rm e}^{-\frac{y}{2\,l_{\rm GZK}}} \, {\rm sin}^2
 \left( \frac{\delta y}{2\,l_{\rm GZK}} \right) \, .
\label{large}
\end{eqnarray}
This displays the presence of an eigenstate which rapidly oscillates between the
original
$|\gamma\rangle$ and $|a\rangle$ states. In terms of the original parameters,
the oscillation length is $L_{\rm osc} \sim M/B$ and, not surprisingly,
it is independent of $l_{\rm GZK}$. Thus, $\delta \sim l_{\rm GZK}/L_{\rm osc}$
gives the number of the photon-axion oscillations within the
mean free path $l_{\rm GZK}$. Thus we can view the coupled
system as particles which are photon-like half of the time, and axion-like
half of the time, so that their probability to interact with the
background photons is one-half of the probability of the decoupled photons
$|\gamma_\perp\rangle$. This explains why
the mean free path of this state is $2 l_{\rm GZK}$.\footnote{In the case of a
magnetic field randomly oriented along different domains and of a strong
mixing $\delta$, we find numerically a mean free path of $3 l_{\rm GZK}/2$,
showing that the quanta
are equally ''shared'' between the axion and the two photon polarizations
states.}

\begin{figure}
\centerline{\includegraphics[width=0.45\hsize,angle=-90]{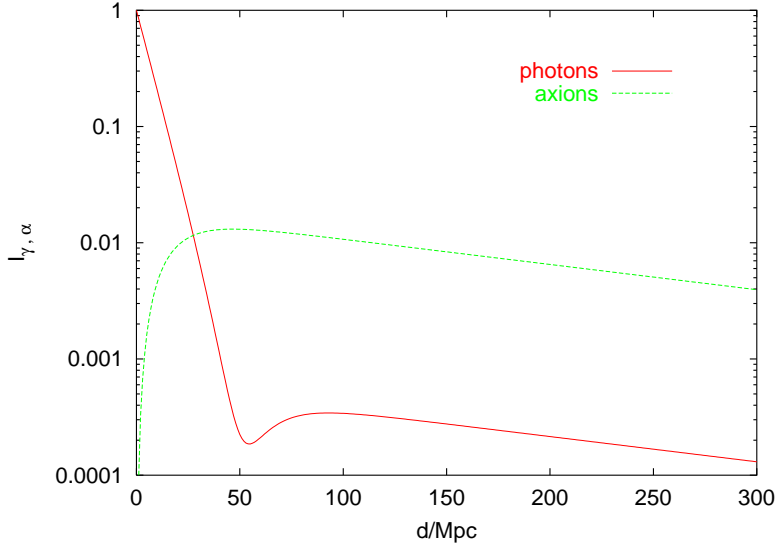}}
\caption{Intensity in photons (initial normalization
$I_\gamma =1$)
and axions traveling
along a constant magnetic field over a distance of  $300\,$ Mpc.}
\label{fig:fig1}
\end{figure}

In the case of small mixing, $\delta \ll 1$, eq.~(\ref{ana1})
reduces to
\begin{eqnarray}
I_\gamma &\simeq& \frac{1}{2} \, {\rm e}^{-\frac{y}{l_{\rm GZK}}} +
\frac{1}{2 \left(1-4 \delta^2 \right)} \left[ \left(1-\delta^2 \right)
{\rm e}^{-\frac{1-\delta^2}{2 l_{\rm GZK}}y} -\delta^2
{\rm e}^{-\frac{\delta^2 y}{2\,l_{\rm GZK}}}
\right]^2 \nonumber\\
I_a &\simeq& \frac{\delta^2}{2} \left[{\rm e}^{-\frac{y}{2\,l_{\rm GZK}}}-
{\rm e}^{-\frac{\delta^2 y}{2 l_{\rm GZK}}}
\right]^2 \, .
\label{small}
\end{eqnarray}
This shows the presence of a long-lived mode,
characterized by the mean free path $l_{\rm GZK} / \delta^2 \gg l_{\rm GZK}$.
To see how it arises, we consider a simple example depicted in Fig.~\ref{fig:fig1},
where we plot $I_{\gamma,a}$ as functions of $y$, taking the
parameters: $l_{\rm GZK} = 6.4 {\rm Mpc}, \, B = 1 {\rm nG}, \,
M=4 \cdot 10^{11}$ GeV (this value of the photon-axion coupling
is close to its experimental bound \cite{bound})
 corresponding to $\delta \simeq 0.18$. These are also the values of
$B$ and $M$ that have been used in~\cite{ckt}.
The photon intensity $I_\gamma$ initially decreases\footnote{We
neglect the secondary photons formed by the scattering of the primaries
against the background photons.} as ${\rm exp} \left( - y / l_{\rm GZK}
\right)\,$.
Meanwhile, some photons convert into axions, and the intensity $I_a$ (initially
taken to be zero\footnote{This is a conservative assumption, since if
initial axions
are also produced
at the source, then our mechanism will still work, the only difference being
that the required
sources can have lower intensities.}) increases. For a small mixing, the
quantity $\delta \sim l_{\rm GZK}/L_{\rm osc}$
gives the amplitude for the process $|\gamma\rangle \rightarrow |a\rangle$ to
occur
within the mean free path of the photons. Thus, $\delta^2$ is
approximately the fraction of the original photons which
converts into axions before being depleted by the interaction with the
background photons. This is confirmed by Eq.~(\ref{small})
and by the numerical evolution shown, which indicate that the axion
intensity peaks at $\sim \delta^2$. From this point on, $I_a > I_\gamma$
and the subsequent photon-axion oscillations lead to a decrease of $I_a$.
The number of axions decreases by a fraction $\delta^2$ per each distance
$\Delta y = l_{\rm GZK}$ traveled. To leading order in $\delta^2$,
this effect is not compensated by those photons which convert back into axions,
since in the case of small mixing the photons in the
beam are more likely to be dissipated by the background photons
than oscillate back into axions. Thus, $I_a \propto I_{a,max} \left( 1 -
\delta^2 \right)^{y/l_{\rm GZK}} \sim
\delta^2 {\rm exp} \left( - \delta^2 y/l_{\rm GZK} \right)$.

This argument explains the presence of an
eigenstate with the mean free path
$l_{\rm GZK} /\delta^2 \gg l_{\rm GZK}$ in Eq.~(\ref{small}).
The axion intensity in this mode relative to the total initial photon intensity
is small, of order $\delta^2$, due to the small
probability for $|\gamma\rangle
\rightarrow |a\rangle$ oscillations. The photon-like component of this mode is
suppressed by a further power of $\delta^2$, because these photons arise from
the double conversion $|\gamma\rangle \rightarrow |a\rangle \rightarrow
|\gamma\rangle$. Most of these photons disappear quickly,
within a distance $y \sim l_{\rm GZK}$ because of the interaction with
background photons. However, the incessant conversion of the surviving
axions in the beam back into photons provides for a long-lasting source of
photons, yielding a slowly decreasing photon
intensity $I_\gamma \sim \delta^4 {\rm exp} \left( - \delta^2 y/l_{\rm GZK}
\right)$. The rise of this ``second population"
of photons is clearly visible in Fig.~\ref{fig:fig1}. The distance $y_*$ at
which these photons reach the level of the surviving intensity of the primary
photons in the beam which never underwent any interactions is roughly given by
$e^{-y_*/l_{GZK}} \simeq \delta^4$. In the present example, $y_*
\simeq 45 $ Mpc, in good agreement with the value shown in the figure.

\section{Consequences for the UHECR}
\setcounter{footnote}{0}

Until now we have considered the evolution of the photon-axion system in a
constant magnetic field. However, the intergalactic magnetic field is
unlikely to be completely uniform. Instead, it is more reasonable
to take the magnetic fields with a strength close to the observed upper bound
of few $\times 10^{-9}$ G to be homogeneous within domains
of a typical size $\sim$ Mpc \cite{magn}, but with their orientation
randomly varying from domain to domain (the same assumptions
were made in \cite{ckt}). We will now look at the
evolution of the photon-axion beam in such a universe.
Photons with polarization perpendicular to ${\vec B}$ do not couple to
axions, so the photon-axion mixing involves photons whose helicity changes
from domain to domain \cite{osc}. If it were
not for the interaction with the background photons, the quanta in the beam would
be roughly equipartitioned between
the axion component and the two photon polarizations after traversing
several domains. This is what is expected
to happen to the visible
photons released in SNe explosions, as explained in
\cite{ckt}. For the range of energies we are interested in, the
interaction with the background photons cannot be neglected, and so the depletion
effect described in the previous section must be taken into account.

It is easy to extend the formulae from the previous section to the general
situation. If the beam is still taken to travel along the $y$ direction, it
can be  described by a vector in the basis $\left( \vert \gamma_x \rangle ,\,
\vert \gamma_z \rangle ,\, \vert a \rangle \right)\,$, where  $\vert
\gamma_{\left\{ x,z \right\}} \rangle$ denote photons polarized along the
$\left\{ x,z \right\}$ axis. Inside each domain, the transfer matrix is such
that it can be rotated to the $ 2 \times 2$ matrix~(\ref{sol2}) for the
coupled axion-photon components, plus a third term ${\rm exp } \left(
- y / 2 \, l_{\rm GZK} \right)$ describing the evolution of the decoupled
photon polarization orthogonal to ${\vec B}\,$.

The resulting transfer equation is:
\begin{eqnarray}
\left( \begin{array}{c} \gamma_x \\ \gamma_z \\ a
\end{array} \right)
= {\rm e}^{i\,E\,y} \, \left[ \, T_0 \, {\rm e}^{\lambda_0\,y}
+T_1 \, {\rm e}^{\lambda_1\,y} + T_2 \, {\rm e}^{\lambda_2\,y} \, \right] \,
\left( \begin{array}{c}
\gamma_x \\ \gamma_z \\ a
\end{array} \right)_0
\end{eqnarray}
where $\lambda_{1,2}$ have been given above,
\begin{equation}
\lambda_0 \equiv -\,\frac{1}{2\,l_{\rm GZK}} \,\,,
\end{equation}
and the three transfer matrices are
\begin{eqnarray}
T_0 &\equiv& \left( \begin{array}{ccc}
{\rm sin}^2 \theta & -\, {\rm cos} \theta \, {\rm sin} \theta & 0 \\
-\, {\rm cos} \theta \, {\rm sin} \theta & {\rm cos}^2 \theta & 0 \\
0 & 0 & 0
\end{array} \right) \,\,, \\
T_1 &\equiv& \left( \begin{array}{ccc}
\frac{1+\sqrt{1-4\,\delta^2}}{2\,\sqrt{1-4\,\delta^2}}
\, {\rm cos}^2 \theta &
\frac{1+\sqrt{1-4\,\delta^2}}{2\,\sqrt{1-4\,\delta^2}}
\, {\rm cos} \theta \, {\rm sin} \theta &
-\,\frac{\delta}{\sqrt{1-4\,\delta^2}} \,{\rm cos} \theta \\
 \frac{1+\sqrt{1-4\,\delta^2}}{2\,\sqrt{1-4\,\delta^2}}
 \, {\rm cos} \theta \, {\rm sin} \theta &
\frac{1+\sqrt{1-4\,\delta^2}}{2\,\sqrt{1-4\,\delta^2}} \, {\rm sin}^2 \theta &
-\,\frac{\delta}{\sqrt{1-4\,\delta^2}} \,{\rm sin} \theta \\
\frac{\delta}{\sqrt{1-4\,\delta^2}} \,{\rm cos} \theta &
\frac{\delta}{\sqrt{1-4\,\delta^2}} \,{\rm sin} \theta &
-\,\frac{1-\sqrt{1-4\,\delta^2}}{2\,\sqrt{1-4\,\delta^2}}
\end{array} \right) \\
T_2 &\equiv& \left( \begin{array}{ccc}
-\,\frac{1-\sqrt{1-4\,\delta^2}}{2\,\sqrt{1-4\,\delta^2}} \, {\rm cos}^2 \theta &
-\,\frac{1-\sqrt{1-4\,\delta^2}}{2\,\sqrt{1-4\,\delta^2}}
\, {\rm cos} \theta \, {\rm sin} \theta &
\frac{\delta}{\sqrt{1-4\,\delta^2}} \,{\rm cos} \theta \\
-\,\frac{1-\sqrt{1-4\,\delta^2}}{2\,\sqrt{1-4\,\delta^2}}
\, {\rm cos} \theta \, {\rm sin} \theta &
-\,\frac{1-\sqrt{1-4\,\delta^2}}{2\,\sqrt{1-4\,\delta^2}}
\, {\rm sin}^2 \theta &
\frac{\delta}{\sqrt{1-4\,\delta^2}} \,{\rm sin} \theta \\
-\,\frac{\delta}{\sqrt{1-4\,\delta^2}} \,{\rm cos} \theta &
-\,\frac{\delta}{\sqrt{1-4\,\delta^2}} \,{\rm sin} \theta &
\frac{1+\sqrt{1-4\,\delta^2}}{2\,\sqrt{1-4\,\delta^2}}
\end{array} \right) ~,
\end{eqnarray}
with $\theta$ being the angle between the magnetic field and the $x$ axis.
The overall transfer matrix
is then simply the product of the transfer matrices in each domain.

\begin{figure}
\centerline{\includegraphics[width=0.45\hsize,angle=-90]{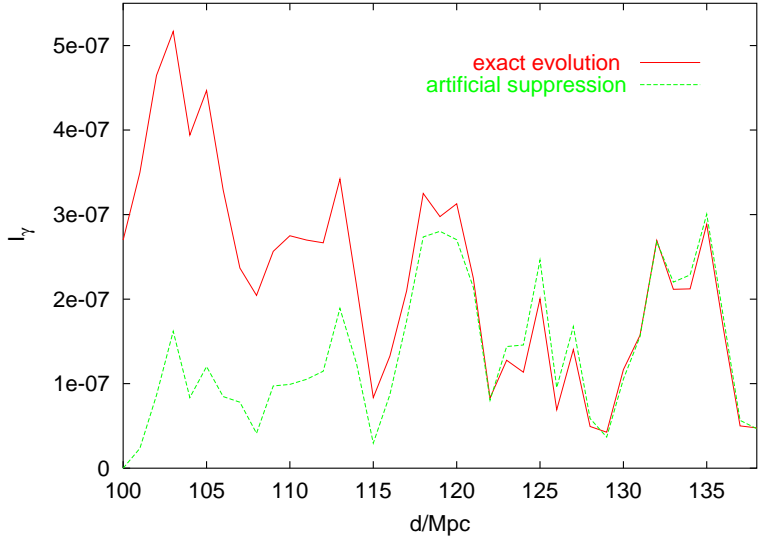}}
\caption{Evolution of the photon intensity across several domains with coherent
magnetic field of length $1$ Mpc each, for $l_{\rm GZK} = 6.4$ Mpc. The
horizontal axis indicates the distance from the source. The exact evolution is
compared to the case in which the intensity of photons is (artificially) set
to zero at $d=100$ Mpc. The two evolutions converge within a few $l_{\rm GZK}$
distances.}
\label{fig:fig2}
\end{figure}

The evolution of the system is qualitatively similar to the one for a constant
magnetic field illustrated in Fig.~\ref{fig:fig1}. However, the interference
effects modify the quantitative results whenever the domain size is
comparable to, or smaller than, the absorption length $l_{\rm GZK}\,$. For
example, consider the evolution of the photon intensity illustrated
 in Fig.~\ref{fig:fig2}. The curve denoted ``exact evolution'' describes
$I_\gamma$ as a function of the distance from the source. All parameters are
as in Fig.~\ref{fig:fig1} except that ${\vec B}$ is randomly oriented in
each domain. The random orientation of $\vec B$
is responsible for the stochastic character of $I_\gamma$ seen in the figure.
At these distances from the source the intensity in axions $I_a$ is much
greater than the one in photons,  decreasing smoothly as in
Fig.~\ref{fig:fig1}. The overall intensity $I_\gamma+I_a$ is thus also smoothly
decreasing. By comparing the value of $I_\gamma$ of the two figures (recalling that
in both cases the photon intensity is normalized to unity at the source), one
notices that $I_\gamma$ is more suppressed by a randomly pointing ${\vec B}\,$.
This effect might be
interpreted as a decrease of the effective mixing parameter $\delta$.
As a consequence, fewer photons will convert into axions before being depleted
by the interaction with the background photons. However, for the very same reason,
the decrease of $I_\gamma$ at distances $\gg l_{\rm GZK}$ appears
to be much milder in the numerical simulations
than the one computed for a constant magnetic field.

An interesting consequence for the transmission of UHECRs is that for
sufficiently distant sources the
only photons in the beam are the ones which have been generated by the axions
within the previous $\sim 4 - 5 \, l_{\rm GZK}\,$
distances. Otherwise they would have been depleted by the
interactions with the background photons.\footnote{We remark
that this concerns the {\it primary} photons, i.e.
the ones which have not interacted with the background photons.}
Since within each domain only the polarization along the magnetic
field is actually generated, the UHECRs reaching the Earth
would be very sensitive to the orientation of ${\vec B}$ in domains within the
sphere of radius $\sim 5 \, l_{\rm GZK}\,$ from us. Consequently the
incoming photons should have slightly different
polarizations along different lines of sight. The magnitude
of this effect depends on the relative size of the domains and on the value
of $l_{\rm GZK}$. If the beam crosses
several domains in the last few $l_{\rm GZK}$ distances, the
differences in the polarization may average out.

We also note that the presence of domains of ${\vec B}$ with different orientation
along different lines of sight can lead to differences in the
intensities of UHECRs from different directions, even if the
distribution of sources were homogeneous and isotropic. This is because
only the magnetic field perpendicular to the line of
travel contributes to the photon-axion mixing in each domain. Thus, we expect lower
fluxes from regions in which the magnetic field is mostly oriented in the
direction pointing toward the Earth. At present there is not enough
experimental data about UHECRs and extragalactic magnetic fields
to enable us to test these predictions. However, this is  a
very distinctive signature that may eventually be tested experimentally.
We illustrate these effects in Fig.~\ref{fig:fig2}, where we have artificially
removed the photons in the beam at a distance $d=100$ Mpc from the source, and
compared the subsequent evolution to the exact one, without this
artificial suppression. One can easily see that the two curves approach each
other within a few $l_{\rm GZK}$ distances.
Since in this case the evolution is stochastic, the curves themselves
behave very differently according to different random choices
for the orientations of ${\vec B} \,$. However, in
all the different simulations the two curves approach each other
very quickly, irrespectively of the particulars of the magnetic
configuration.

\begin{figure}
\centerline{\includegraphics[width=0.45\hsize]{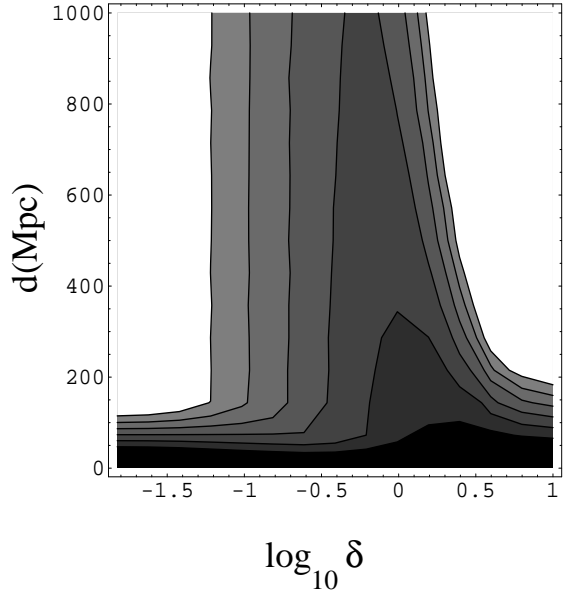}}
\caption{Intensity of photons $I_\gamma$ (logarithmic units) from a single
source at a distance $d$ (given in Mpc on the vertical axis) for an arbitrary
value of $\delta$ (${\rm Log}_{10} \delta$ shown on the horizontal axis).
The intensity at the source is normalized to unity and different contour
lines are at $I_\gamma = 10^{-3} \,, 10^{-4} \,, \dots \,, 10^{-8}\,$. Darker
regions correspond to greater intensity.}
\label{fig:fig3}
\end{figure}

In Fig.~\ref{fig:fig2} the mild decrease of the photon
intensity at large distances is hidden by its stochastic behavior. To see the
decrease, one has to average the evolution over several configurations of
${\vec B}\,$. This averaging procedure is appropriate if we discuss
UHECRs independently of their arrival directions, which have crossed
different, uncorrelated configurations of the magnetic field. The remaining
figures that we show have been obtained using this averaging procedure.

In Fig.~\ref{fig:fig3} we display the evolution of the photon
intensity as a function of the distance from the source (vertical
axis) and of the mixing parameter $\delta\,$ (horizontal axis). As
before $l_{\rm GZK}=6.4$ Mpc. The photon intensity drops rapidly
both at low and high values of $\delta\,$. We note that the
surviving photon intensity is maximized for $\delta$ between $\sim
0.1$ and order unity. This gives the region of preferred values of
the photon-axion coupling parameter $M$ and the magnetic field $B$
where our mechanism is efficient (see Eq.~(\ref{deltaref})).
Interestingly, the choice of parameters adopted in~\cite{ckt} to
explain the SNe dimming in terms of the photon-axion conversion is
precisely within this region.

\begin{figure}
\centerline{\includegraphics[width=0.45\hsize,angle=-90]{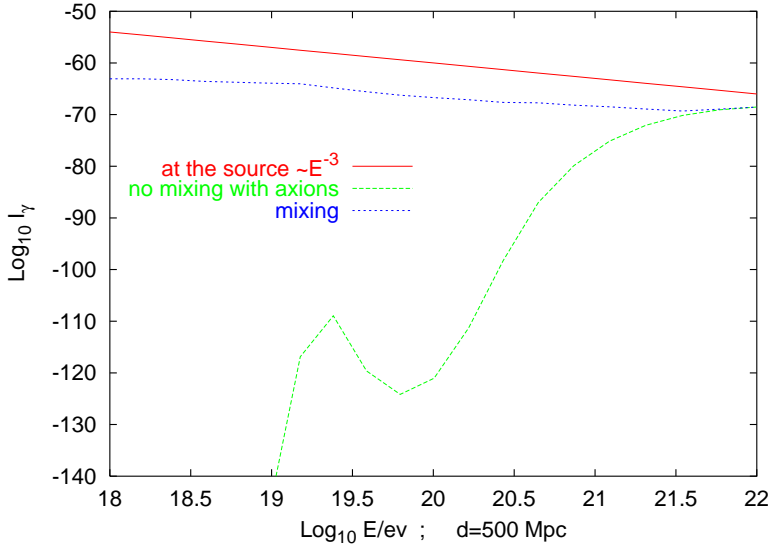}}
\caption{Photons from a source at $d=500\,$ Mpc from the Earth. The initial
spectrum is assumed to have an energy dependence $I_\gamma \propto E^{-\,3}$.
The spectrum of primary
photons
reaching the Earth is shown both with and without the photon-axion mixing.
See the main text for details.}
\label{fig:fig4}
\end{figure}

Ultimately we would like to find how the photon-axion mixing
affects the spectrum of UHECRs observed on Earth. A precise
computation, which requires the inclusion of the secondary
particles produced  when a photon in the beam interacts with a
background photon, is beyond the scope of this paper. However, a
rough but illustrative estimate can be obtained relatively simply.
We begin by considering an explicit example of the evolution of a
primary photon spectrum from a single source at $d=500\,$ Mpc from
the Earth, shown in Fig~\ref{fig:fig4}. The intensity at the
source is assumed to scale as $E^{-3}$ (the overall normalization
is irrelevant for the present discussion). Even without the mixing
with axions the spectrum of primary photons reaching the Earth is
strongly modified due to their interaction with the background
photons. The reason is that the mean free path $l_{\rm GZK}$ is
strongly dependent on their energy. For the energy range shown in
Fig.~\ref{fig:fig4}, it ranges from $\sim 0.3$ Mpc for $E =
10^{18}$ eV up to $\sim 100$ Mpc for $E = 10^{22}$ eV (we have
used the results of~\cite{prjo} summarized in~\cite{aprs}, see
these references for details). The strong increase of $l_{\rm
GZK}$ in this interval explains why the second spectrum in the
figure is peaked at high energies, although the original
spectrum\footnote{The local maximum at $E \sim 2 \cdot 10^{19}$ eV
corresponds to a local maximum of $l_{\rm GZK}$, due to the fact
that at smaller energies the incoming photons are mostly affected
by CMB photons, while at higher energies they are most affected by
radio photons. See~\cite{prjo,aprs} for details.} scaled as
$E^{-3}\,$.The curves shown in Fig.~\ref{fig:fig4} confirm a much
lower depletion of primary photons when the mixing with axions is
taken into account (as in all the previous cases, the axion-photon
mass parameter is taken to be $M=4 \cdot 10^{11}$ GeV).

\begin{figure}
\centerline{\includegraphics[width=0.45\hsize,angle=-90]{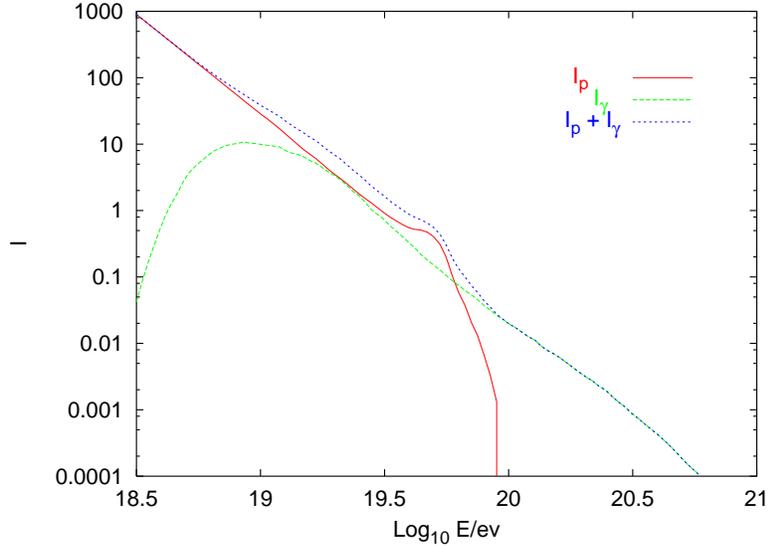}}
\caption{Combined spectra of high energy protons and primary photons, for a
homogeneous and isotropic distribution of sources at distances between $100$
and $1000$ Mpc from the Earth. An energy dependence $I \propto E^{-\,3}$ is
assumed at the source. See the main text for details.}
\label{fig:fig5}
\end{figure}

We finally relate this decreased depletion of
$I_\gamma$ to an apparent absence of a GZK cutoff in the UHECR spectrum. In
Fig.~\ref{fig:fig5} we combined a spectrum of protons of extragalactic origin
with the spectrum of primary photons. Both spectra have been computed for an
isotropic and homogeneous distribution of sources at distances $100 \,
{\rm Mpc} < d < 1000 \, {\rm Mpc}$ from the Earth. For both populations,
the intensity at the source is again taken to have a $I \propto E^{-3}$
dependence. Notice that the proton spectrum $I_p$ exhibits the GZK
cut-off,\footnote{For the proton evolution, we have followed the
results of~\cite{ades} summarized in~\cite{aprs}.}  while -- due to their
coupling with the axions -- this is not the case for the photon spectrum
$I_\gamma\,$. By an appropriate relative choice of $I_p$ and $I_\gamma$ at
the source, it is thus possible to match the two spectra such that the
photon flux ``catches up'' precisely where the proton flux exhibits the GZK
cutoff. The overall spectrum can thus be extended at energies well above
$E_{\rm GZK}\,$.

Before concluding, we stress the limitations of the computations summarized
 in the last two figures. The main one is related to the fact that $I_\gamma$
does not include the secondary photons arising when the photons in
the beam scatter against the background photons. This will lead to
a significant modification of $I_\gamma$. Fig.~\ref{fig:fig5} has
been obtained for a ratio $I_\gamma / I_p = 25,000$ at the source.
This would amount to a flux of about $10^{-23} \, \left( m^2 \, sr
\, s \, GeV \right)^{-1}$ on Earth, if the photons were travelling
unimpeded. We expect that a smaller ratio will be required once
secondary photons are taken into account. Secondly, in all of the
above computations we have neglected the decrease of the energy
(with a consequent change of $l_{\rm GZK}$) of the quanta in the
beam due to the cosmological redshift (for sources at $d=1000$
Mpc, this correspond to about a $30 \%$ decrease of the energy).
Thus, the computation summarized in Fig.~\ref{fig:fig5} should be
understood only as an order of magnitude estimate and as an
illustration of the effect which can be expected. The significance
of this estimate will be highly dependent on which sources for the
photons one is assuming. The problem of accelerating particles at
such energies in different astrophysical environments is completely
independent from the problem of their propagation across intergalactic
media. Thus it is sensible to separate these problems from each other.
Therefore in the present paper we have focused only on the latter,
the propagation problem.

We close this section with a note on the nature of the UHECRs.
Once UHECRs hit the atmosphere of the Earth, they give rise to
extensive air showers. Different incident particles generate
distinct longitudinal profiles, that is the amount of particles as
a function of the amount of atmosphere penetrated by the cascade.
Several analyses, for example the one reported by the Fly's Eye
Collaboration in~\cite{eye}, suggest a change from an iron
dominated composition of the incoming particles at $ \sim 10^{17}$
eV to a proton dominated composition near $10^{19}$ eV.  At higher
energies,  the current statistics are too poor for a conclusive
claim, and one still cannot exclude photons as a relevant fraction
of UHECRs above the GZK cut-off \cite{ave,kkst,aprs}. The most
accurate method to determine the nature of UHECR of such energies
is the study of the muon content of inclined showers (that is, the
ones with incident zenith angle $\theta > 60^0\,$). The strongest
bounds quoted in the literature \cite{aprs} have been obtained in
\cite{ave}, where inclined air showers recorded by the Haverah
Park \cite{haverah} detector have been analyzed. The results of
\cite{ave} are affected by uncertainties in the parameterization
of the flux of cosmic rays. Assuming the parameterization given in
\cite{nawa}, the authors of \cite{ave} obtain that less than $48
\%$ of the observed events above $10^{19} \,$eV can be photons
with a $95 \%$ confidence level. For energies higher than $4
\times 10^{19} \,$eV, they find instead that the content of
photons has to be less than $50 \%$ of the overall flux. Different
results are obtained based on the parameterization given in the
more recent work \cite{sww}. The two upper bounds change to $25
\%$ for energies above $10^{19} \,$eV, while only to $70 \%$ for
energies greater than $4 \times 10^{19} \,$eV \cite{ave}.

Hence, we see that a mixed composition of proton and photons in
the UHECR above the GZK cutoff is compatible with the present
experimental limits. Thus the mechanism we propose here is in
agreement with the data at the moment: it relies precisely on the
assumption that both protons and photons give a sizeable
contribution to the UHECR at energies around $10^{19} \,$eV. The
present experimental results do not yield clear-cut constraints at
higher energies, as manifested by the discrepancy between the
analysis based on the different parameterizations of \cite{nawa}
and \cite{sww}. On the contrary, the higher statistics and better
energy determination expected in the forthcoming Auger Observatory
\cite{auger} have the power of greatly improving the present
bounds. Therefore it may either discover photon UHECRs or yield
significantly stronger bounds on the dynamics of photon-axion
mixing we employ here. It will thus be of interest to subject the
mechanism we propose to more scrutiny.

One more concern about photons being UHECRs could be that most of
the photons that do interact with the background (the secondary
photons) would cascade down to GeV energies and contribute to the
diffuse gamma ray background. If the survival probability of the
UHE photons were too low, then the model would have predicted too
many secondary photons and thus a gamma ray background that is too
large. The average survival probability from Fig.~\ref{fig:fig3}
for the optimal values of $\delta$ are of order $10^{-4}$. The
ratio of total energies contained in the 1 GeV gamma ray
background to the energies contained in UHECRs at $10^{20}$ eV can
be obtained from comparing the measured EGRET flux~\cite{EGRET} to
the various measurements of UHECR fluxes. This ratio turns out to
be of order $10^{-3} - 10^{-4}$, depending on whether a high or
low number for the UHECR flux is adopted. The lower value seems to
fit in our model very well, while for the higher value there could
seemingly be a marginal conflict (of at most an order of
magnitude) with the gamma ray background bounds. However, the
ratio of primary to secondary photons can be higher than the
survival probabilities in Fig.~\ref{fig:fig3} for various
different reasons. First, there could be an initial axion
component in the beam which would dramatically change the
prediction for the ratio of primary to secondary photons. Note
that if the initial beam contained an order unity fraction of the
energy in axions, the primary photon to secondary photon ratio
would also be of order unity. Thus it is clear that even a
relatively small admixture of axions in the initial beam can
drastically improve the primary to secondary photon ratio. Second,
the conversion rate within our galaxy is much larger since the
galactic magnetic field is $\sim 10^3$ times bigger than the
assumed value of the intergalactic magnetic field. If a somewhat
larger value of magnetic field persists for a few kiloparsecs,
then the survival probabilities will be increased by a factor of
few, and this on its own could be sufficient to bring down the
number of secondary photons. Lastly, as we have stressed above,
only a fraction of UHECRs may be photons. Then the total number of
secondary photons would fall below the diffuse gamma ray bounds
even if the primary to secondary photon ratio stays the same. In
this way the signatures of our mechanism would still persist
without any violations of the diffuse gamma ray background bounds.

Finally, a photon component of UHECRs may be related to the
modification of their spectrum by the interaction with the
background photons, once the secondary photons are also taken into
account in the propagation codes. After scattering with the
background photons, an incoming photon of energy ${\cal E}_0$
produces secondary photons which accumulate at energies $< {\cal
E}_0\,$. Starting with some flux at the source and normalizing the
final spectrum to the observed amount of events above the GZK
cutoff, one would typically find an excess of events at lower
energies \cite{prjo}. However here the mixing with the axions can
significantly modify the final spectrum precisely in the direction
of ``balancing out'' this excess at lower energies. Indeed, the
energy of the photons continuously supplied by the axions is given
by the initial energy at the source decreased only by the
expansion of the Universe, and it is greater than that of the
secondary photons present in the beam. A detailed computation of
the final spectrum, including both primary and secondary photons
will be an important test of the viability of the mechanism
proposed here.

\section{Conclusions}
\setcounter{footnote}{0}

An interesting property of axions, which arise in many extensions of
the SM, is that they mix with photons in an external magnetic
field. Photon-axion conversions in extragalactic magnetic fields can
have relevant consequences for astrophysics and cosmology. We have
shown that the same parameters (axion mass, coupling scale, magnetic
field configurations) which provide an alternative explanation for the
dimming of distant SNe \cite{ckt} could also imply an enhancement of
the flux of super-GZK photons arriving to the Earth from faraway
sources. Whether this enhancement is observable clearly depends on the
distribution of the sources and on the intensity of the high-energy
photon spectrum they emit. We have seen here that, for realistic
assumptions, super-GZK photons can be detectable, and may provide an
explanation for the origin of super-GZK events in the UHECR spectrum,
if the required sources of very high energy photons exist. The
viability of this mechanism can be tested by improved simulations of
the photon propagation, and ultimately by the increase of UHECR data
expected over the next few years.

\section*{Acknowledgments}
We thank J. Arons, J. Feng, H.P. Nilles, and G. Raffelt for useful
discussions, and to Dimitry Semikoz for useful comments on the
manuscript. C.C. is supported in part by the DOE OJI grant
DE-FG02-01ER41206 and in part by the NSF grant PHY-0139738. N.K.
is supported in part by a Research Innovation Award from the
Research Corporation. J.T. is supported by the U.S. Department of
Energy under contract W-7405-ENG-36.

\end{document}